\documentclass{aa}  

\usepackage{graphicx}
\usepackage{rotating}
\usepackage{pdflscape}
\usepackage{color}
\usepackage{natbib}
\usepackage{amssymb}
\usepackage{txfonts}
\def\rgc{R$_{\rm gc}$}
\def\kms{km s$^{-1}$}

\begin{document} 

   \title{The Gaia-ESO Survey: Properties of the intermediate age open cluster NGC 4815\thanks{Based on observations made with the ESO/VLT, at Paranal Observatory, under program 188.B-3002 (The Gaia-ESO Public Spectroscopic Survey).}
}
   \titlerunning{Abundance analysis of NGC 4815}

   \author{E. D. Friel\inst{1}, P. Donati\inst{2,3}, A. Bragaglia\inst{3}, H. R. Jacobson\inst{4}, L. Magrini\inst{5}, L. Prisinzano\inst{6}, S. Randich\inst{5}, M. Tosi\inst{2}, T. Cantat-Gaudin\inst{7,8}, A. Vallenari\inst{8}, R. Smiljanic\inst{9,10}, G. Carraro\inst{10}, R. Sordo\inst{8},  E. Maiorca\inst{5}, G. Tautvai\v{s}ien\.{e}\inst{11}, P. Sestito\inst{12}, S. Zaggia\inst{8}, F. M. Jim\'{e}nez-Esteban\inst{13,14}, G. Gilmore\inst{15}, R.~D. Jeffries\inst{16},     E. Alfaro\inst{17},   T. Bensby\inst{18}, S.~E. Koposov\inst{15, 19}, A.~J. Korn\inst{20}, E. Pancino\inst{3,21},  A. Recio-Blanco\inst{22}, 
E. Franciosini\inst{5}, V. Hill\inst{22}, R.~J. Jackson\inst{16},  
  P. de Laverny\inst{22}, L. Morbidelli\inst{5}, G.~G. Sacco\inst{5}, C.~C. Worley\inst{15}, A. Hourihane\inst{15}, M.~T. Costado\inst{17}, P. Jofr\'e\inst{15}, K. Lind\inst{15} 
   }

   \institute{
   Department of Astronomy, Indiana University, Bloomington, IN, USA;   \email{efriel@indiana.edu}
     \and
   Dipartimento di Fisica e Astronomia,  Universit\`a di Bologna, Via Ranzani, 1, 40127, Bologna, Italy
   \and
   INAF-- Osservatorio Astronomico di Bologna, Via Ranzani, 1, 40127, Bologna, Italy
    \and
   Massachusetts Institute of Technology, Kavli Institute of Astrophysics \& Space Research, Cambridge, MA, USA
   \and
   INAF--Osservatorio Astrofisico di Arcetri, Largo E. Fermi, 5, I-50125, Firenze, Italy
   \and
   INAF-- Osservatorio Astronomico di Palermo, Piazza del Parlamento, 1, 90134, Palermo, Italy
   \and
   Dipartimento di Fisica e Astronomia, Universit\`a di  Padova, vicolo Osservatorio 3, 35122 Padova, Italy
   \and 
   INAF- Osservatorio Astronomico di Padova, vicolo Osservatorio 5, 35122 Padova, Italy
   \and 
   Department for Astrophysics, Nicolaus Copernicus Astronomical Center, ul. Rabia\'nska 8, 87-100 Toru\'n, Poland
   \and
   European Southern Observatory, Karl-Schwarschild-Str 2, 85748 Garching bei M\"unchen, Germany
   \and
   Institute of Theoretical Physics and Astronomy, Vilnius University, A. Gostauto 12, 01108 Vilnius, Lithuania
   \and
   Universidad Complutense de Madrid, Plaza de la Ciencia 3, 28040 Madrid, Spain
   \and
   Centro de Astrobiolog\'{\i}a (INTA-CSIC), Departamento de Astrof\'{\i}sica, PO Box 78, E-28691, Villanueva de la Ca\~nada, Madrid, Spain
   \and
      Suffolk University, Madrid Campus, C/ Valle de la Vi–a 3, 28003, Madrid, Spain
   \and
Institute of Astronomy, University of Cambridge, Madingley Road, Cambridge CB3 0HA, United Kingdom
\and
Astrophysics Group, Research Institute for the Environment, Physical Sciences and Applied Mathematics, Keele University, Keele, Staffordshire ST5 5BG, United Kingdom
   \and
 Instituto de Astrof\'{i}sica de Andaluc\'{i}a-CSIC, Apdo. 3004, 18080, Granada, Spain
\and
   Lund Observatory, Department of Astronomy and Theoretical Physics, Box 43, SE-221 00 Lund, Sweden 
   \and
Moscow M.V. Lomonosov State University, Sternberg Astronomical Institute, Universitetskij pr., 13, 119992 Moscow, Russia
\and
Department of Physics and Astronomy, Uppsala University, Box 516, SE-75120 Uppsala, Sweden
\and
ASI Science Data Center, Via del Politecnico SNC, 00133 Roma, Italy
 \and
Laboratoire Lagrange (UMR7293), Universit\'e de Nice Sophia Antipolis, CNRS,Observatoire de la C\^ote d'Azur, BP 4229,F-06304 Nice cedex 4, France
}
   
   \authorrunning{Friel et al.}
 \date{Received / accepted}

 
  \abstract
   { NGC 4815 is a populous $\sim$ 500 Myr open cluster at \rgc $\sim$ 7 kpc observed in the first six months of the Gaia-ESO Survey. Located in the inner Galactic disk, NGC 4815 is an important potential tracer of the abundance gradient, where relatively few intermediate age open clusters are found.}
   {The Gaia-ESO Survey data can provide an improved characterization of the cluster properties, such as  age, distance, reddening, and abundance profile. }
   {We use the survey derived radial velocities, stellar atmospheric parameters, metallicity, and elemental abundances for stars targeted as potential members of this cluster to carry out an analysis of cluster properties.  The radial velocity distribution of stars in the cluster field is used to define the cluster systemic velocity and derive likely cluster membership for stars observed by the Gaia-ESO Survey.  We investigate the distributions of Fe and Fe-peak elements, alpha-elements, and the light elements Na and Al and characterize the cluster's internal chemical homogeneity comparing it to the properties of radial velocity non-member stars.  Utilizing these cluster properties, the cluster color-magnitude diagram is analyzed and  theoretical isochrones are fit to derive cluster reddening, distance, and age.}
   {NGC 4815 is found to have a mean metallicity of [Fe/H]$=+0.03 \pm 0.05$ dex (s.d.).  Elemental abundances of cluster members show typically very small internal variation, with internal dispersions of $\sim$ 0.05 dex.  The alpha-elements [Ca/Fe] and [Si/Fe] show solar ratios, but [Mg/Fe] is moderately enhanced, while [Ti/Fe] appears slightly deficient.  As with many open clusters, the light elements [Na/Fe] and [Al/Fe] are enhanced, [Na/Fe] significantly so, although the role of internal mixing and the assumption of LTE in the analysis remain to be investigated.  From isochrone fits to color-magnitude diagrams, we find a cluster age of 0.5 to 0.63 Gyr, a reddening of $E(B-V) = $ 0.59 to 0.65, and a distance modulus $(m-M)_0 = 11.95$ to 12.20, depending on the choice of theoretical models, leading to a Galactocentric distance of 6.9 kpc.}
  {}

   \keywords{Galaxy: open clusters and associations: individual: NGC 4815 -- Stars: abundances, Hertzsprung-Russell and C-M diagrams}

   \maketitle
   
%

\section{Introduction}

The Gaia-ESO Survey \citep{GES1,GES2} is an ambitious high-resolution public spectroscopic survey of stars in the Galaxy carried out with the FLAMES multi-object spectrograph on the ESO Very Large Telescope.  Over its 5-year lifetime, the survey will target up to 90 to 100 open clusters of a wide range of properties, ages, and locations in the Galaxy (Randich et al. 2014, in prep.).   Survey data are processed and analyzed in a homogeneous way to ensure a final data set of stellar kinematic, atmospheric, and chemical abundance  properties that are derived in a uniform and consistent manner (Lewis et al. 2014, in prep., Sacco et al. 2014, submitted., Smiljanic et al. 2014, in prep.).   Among the many long-term goals of the survey are an improved understanding of the formation and chemical evolution of the Galactic disk as traced by the open cluster population.    

Most stars are born in aggregates of stars, most of which are unbound and dissolve into the field population \citep{lada03}.  Those clusters that remain serve as important indicators that permit the study of a wide variety of issues in cluster formation and Galactic evolution, including investigations of the mass function and survivability of star clusters \citep{larsen09}, the process of  cluster dispersal especially as traced by the chemical tagging of former cluster members \citep{FB2002,BKF2010}, and the development and evolution of the abundance gradients in the Galactic disk \citep{magrini10, jpf}.  Key to addressing these last issues on chemical evolution is the question of the chemical homogeneity of the surviving clusters, and by implication those that have dispersed into the field population.  Careful work by \citet{deSilva06,deSilva07,deSilva11}, for example, has demonstrated the chemical homogeneity in clusters such as the Hyades and Cr261, as well as the potential to trace dispersed cluster members into the surrounding field.  

It remains an open question whether all elements display such uniformity in open clusters, however.  Light elements such as Na may show the effects of internal nucleosynthesis and mixing to the surface in evolved stars \citep{sm12}, although the situation is complicated by the potential influence of non-LTE effects.  Open clusters do not appear to share the common  (anti-)correlations between the light and alpha-elements  seen in the globular cluster population  \citep{carretta10}.  Work by  \citet{deSilva09} and \citet{brag12}, for example, finds no evidence of an  O-Na anti-correlation or the presence of the extreme O-depletions or Na-enhancements see in in the globulars.  The unusual cluster NGC 6791 has prompted much attention for some first signs of globular cluster-like Na-O anti-correlations \citep{Geisler12}, but further studies are not finding evidence for intrinsic abundance dispersions in the cluster \citep{n6791_14}.  Larger and uniform samples of abundances in open clusters offer the potential to distinguish these behaviours and define chemical characteristics that separate cluster populations.  

If open clusters are to trace these disk properties and their evolution, we must first be able to securely define the cluster members and distinguish them from the surrounding field population.  And we must place the cluster properties such as age, abundance, and location in the disk on uniform and consistently determined scales so that details of distributions can be probed with confidence that we are not simply exploring systematic differences between approaches or methodology.  With these basic properties well characterized, clusters become indicators of both their initial environments (through their chemical signatures) and their current environments, allowing the possibility of using them to explore their migration through the Galactic disk.

The Gaia-ESO survey aims to address all of these scientific goals based on  substantial samples of clusters and cluster members and, most importantly, cluster properties and abundances on a uniform and internally consistent scale.  The  observations of the first three intermediate-aged clusters allow us to explore the adequacy of the survey data and establish procedures that will yield a uniform set of cluster parameters.  Progress on the larger Galactic context of the cluster data must await the samples to come, but the initial survey data allow the first steps to be taken.  

The first six months of GES survey observations included three intermediate-age clusters that allow us to begin to explore these issues.  The clusters Trumpler 20, NGC 4815, and NGC 6705 (M11)  differ in many ways, but all are located inside the solar circle.  This location makes them important probes of the abundance gradient in the inner regions of the Galaxy, but also means that their study will be complicated by high and possibly variable reddening, a potentially large and complex field  population, and the contamination of many non-members superimposed on the cluster field.  As a result, the basic properties of these clusters are often poorly constrained.   The Gaia-ESO survey data allow us to define these properties, particularly by comparison to the surrounding field, understanding the chemical profile of these clusters and improving the cluster parameters such as age, distance, and reddening, which are critical to their use in tracing abundance gradients in the disk.

This paper presents the results of data for stars in the field of NGC 4815.   Similar analysis of the other early Gaia-ESO Survey cluster observations are presented in companion papers (Tr 20 in \citealt{pdTr20};  NGC 6705 in Cantat-Gaudin et al. 2014, in prep.). 
Each of these clusters presents a particular characteristic (Tr 20 shows an unusual morphology of the red giant clump; NGC 6705 is unusually massive for an open cluster;  NGC 4815 is embedded in a rich field population) that allows the focus on different aspects of the analysis and interpretation of the GES data in the context of available information from the literature.  The three clusters are considered together in the broader context of chemical evolution in the Galaxy in \citet{magrini14}.

 NGC 4815 is an intermediate age cluster located inside the solar Galactocentric radius, at RA=12:57:59, Dec=$-$64:57:36 and Galactic coordinates 
$\ell = 303.6$, $b = -2.1$.  It shows distinctly against a crowded field at this low latitude.  Located
approximately 2.5 kpc away, the line of sight to the cluster passes through both the Sagittarius and the
Scutum-Centaurus spiral arms, with the cluster located beyond, but close to, the Sagittarius arm.  The cluster has been the subject of several photometric studies, but has not been observed spectroscopically until now.  

The paper is organized as follows:  In Sec. 2 we briefly describe our understanding of the properties of NGC 4815 based on past studies.  Sec. 3 discusses the results from the Gaia-ESO Survey data, beginning with a description of the observed sample, followed by an analysis of the radial velocities and then the stellar abundances.  Sec. 4 utilizes the results on cluster membership and metallicity to derive cluster properties based on fitting of theoretical isochrones.  Finally, Sec. 5 summarizes our results.  

\section{Cluster properties from the literature}
 
The earliest photometry of the cluster was presented by \citet{kf91}, 
who obtained
$UBV$ photometry of some 600 stars in a 2.5$\arcmin$ by 4.2$\arcmin$ field of view to $V \sim$ 19.
They derived an apparent distance modulus of $(m-M)_{V} = 14.75$ and reddening $E(B-V) = 0.78$, which they noted as especially uncertain, yielding an intrinsic distance modulus $(m-M)_{0} = 12.33$ and a distance of 2.9
kpc (assuming $A_{V}= 3.1E(B-V)$).  They also fit a combination of theoretical isochrones to derive a cluster age of 200 Myr.
They pointed out the strong
field star contamination in the cluster field, but noted that in spite of this a clear cluster main-sequence
and turnoff region was visible in the color-magnitude diagram.  A later photometric study in $B$ and $V$ by
\citet{co94} over a much larger 12.3$\arcmin$ by 12.3$\arcmin$ field of view, consisting of 2500 stars to $V \sim$ 19.5, yielded a somewhat shorter 
 distance modulus of $(m-M)_{V} = 14.10$ and reddening consistent within the uncertainties of
$E(B-V)=0.70$, yielding a true distance modulus of $(m-M)_{0} = 11.93$ and a distance of 2.4 kpc.  They also found that the cluster radius is less than about 4.6$\arcmin$, and note a significant contamination by field stars.  The application of a technique relying on synthetic color-magnitude diagrams built up from theoretical isochrones with contributions from photometric errors and binary systems \citep{chiosi89} resulted in a best fit for a slightly
sub-solar metallicity of [Fe/H] = $-0.40$, and a cluster age of 500 Myr.   

\citet{pris01} followed
with a deep $VI$ study to $V \sim$ 25, over a 9$\arcmin$ by 9$\arcmin$ area, intended to investigate the mass function of the cluster.  While they did
not independently determine the cluster parameters, they noted that the CMD of the cluster field was not
strongly distinguished from that of the surrounding fields populated by the Galactic disk.  They also
determined a cluster radius of 3.6$\arcmin$, significantly smaller than that determined by \citet{co94}, but also at the limit of the field size they observed.  They noted that the cluster main-sequence
merges with the field population for $V$ $>$ 21, beyond which it is completely confused with the Galactic
field.  Because of this confusion with the field at the faintest magnitudes, a mass estimate for the cluster is difficult to determine, but considering the mass function for stars more massive than $\sim 0.8 M_{\sun}$, corresponding to this magnitude limit, the cluster mass is estimated to be $880 \pm 230  M_{\sun}$.  

Finally, \citet{sagar01} obtained $BVI$ photometry over a small field of $\sim$2.1$\arcmin$ by 3.3$\arcmin$ centered on the cluster, with an offset position with respect to the cluster center to sample the field population.  Fitting \citet{bert94} isochrones to the data, they derived a reddening $E(B-V) = 0.72$, an intrinsic distance modulus of $(m-M)_{0} = 12.2$, giving a distance of 2.75 kpc, and an age of $\sim$ 400 Myr. 

These studies agree within the uncertainties that NGC 4815 is an intermediate age cluster of $\sim$0.4 - 0.5 Gyr, located at a Galactocentric distance of 6.9 kpc (assuming the solar Galactocentric distance of
8.0 kpc, \citealt{mal13}).  These properties make NGC 4815 a particularly interesting cluster to study the behavior of the disk abundance
gradient in the region inside the solar circle, where there are relatively few old clusters known and
studied (\citealt{magrini10}, Jacobson et al. 2014, in prep.).

\section{Results from the Gaia-ESO Survey data}
\subsection{Gaia-ESO Survey observations}
\begin{figure}
\centering
\includegraphics[width=1.8\hsize]{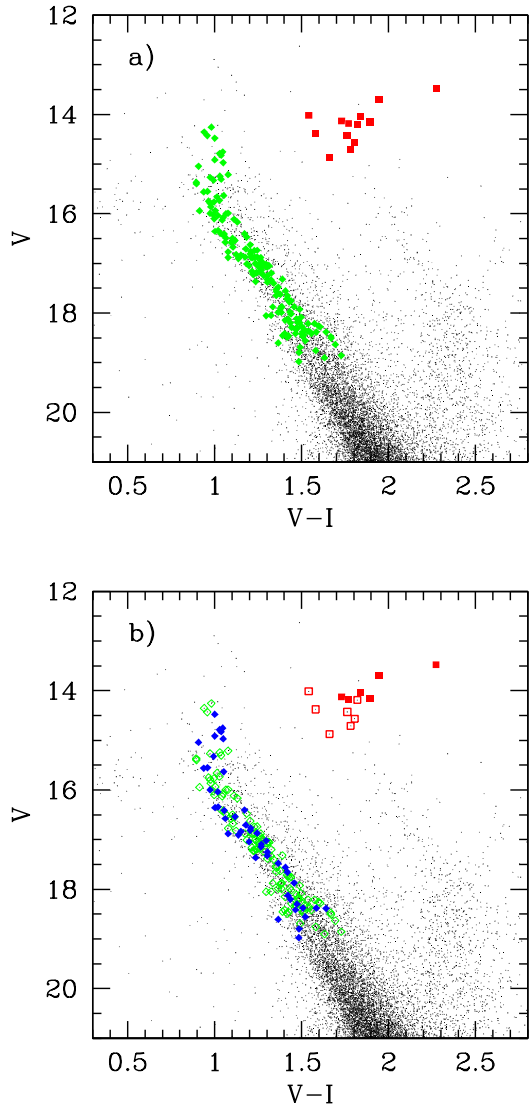}
\caption{$VI$ color-magnitude diagram for NGC 4815 from \citet{pris01} with stars observed by Gaia-ESO Survey indicated.  {\it  Panel a)} Stars observed with UVES indicated with red squares.  GIRAFFE targets are indicated with green diamonds.  An additional 39 targets are not shown as they are lacking $VI$ photometry.   {\it Panel b):} Cluster members on the basis of radial velocity determination are indicated with filled red squares (for UVES targets) and filled blue diamonds (for GIRAFFE targets).  Stars with velocities differing more than 4 \kms~  from the cluster mean are indicated with open symbols. }
\label{cmd_vi}
\end{figure}

The general aims and details of the process for the selection of targets in open clusters in the Gaia-ESO survey is described in Bragaglia et al. (2014, in prep).  In the case of NGC 4815, targets were selected from consideration of the combined photometry from \citet{pris01}, 2MASS, and two unpublished optical studies, in $BVI$ using FORS on the VLT (Bragaglia et al., in prep.), and $BV$
from WFI on the 2.2m (Zaggia et al., in prep).  Targets were selected within an area covering twice the radius of the cluster determined by \citet{pris01}, namely 7.2$\arcmin$.   Targets for UVES observations were limited to stars in the region
of the red clump; 14 of these cluster candidates were observed using the 580nm setup, covering 4760-6840 \AA.  Targets for GIRAFFE observations were
selected to lie along the main-sequence extending from the turnoff region at $V \sim$ 14.2 to 19.  The brighter sample of stars, from $V \sim$ 14.2 to 17, was observed with GIRAFFE grating HR09 (5145-5356 \AA), while those with 16.8 $< V <$ 19 were observed with grating HR15n (6470-6790 \AA).  Twenty-one of the fainter stars in the bright magnitude interval were observed with both setups, resulting in a total sample size of GIRAFFE observations of 204 stars.    The location of those stars with $VI$ photometry is indicated in the color-magnitude diagram from \citet{pris01} in Fig.~\ref{cmd_vi}. One additional giant star and 38 main-sequence stars with only $BV$ photometry were also observed.  

NGC 4815 observations were taken on 5 nights in April and May of 2012.  Observations of the setup using GIRAFFE grating HR09 consisted of three 50-minute plus six 25-minute observing blocks.   Those with the HR15n grating consisted of three 50-minute plus four 25-minute observing blocks.  Signal-to-noise ratios for the final UVES spectra range from 20 to 130, with typical values being $\sim$ 65.  For the GIRAFFE spectra, S/N ratios were typically lower, ranging from 10 to 80, with a median of $\sim$30.  

Data were processed with the Gaia-ESO Survey pipeline as described in Lewis et al. (2014, in prep.) for GIRAFFE data and \citet{sacco14}for UVES data.  Results discussed here are derived from the first internal Data Release (GESviDR1Final) made available to the members of the Gaia-ESO Survey collaboration.  Radial velocities are the recommended values derived from the pipeline processing for both GIRAFFE (see Lewis et al. 2014, in prep.) and UVES observations cited{sacco14}.   For the UVES observations, final velocities are averages of determinations  from the upper and lower spectral regions.  For GIRAFFE observations, when velocities were available from both the HR09 and the HR15n grating setups, those determined from the higher S/N exposure were used as these have significantly smaller velocity errors.  In all cases, these were the HR15n observations, which had S/N values of $\sim$ 50 to 65 versus much lower values of $<$ 15 for HR09 observations.  As a result, velocities for 113 stars came from HR15n observations, while 91  utilized HR09 observations.  

\subsection{Radial velocity analysis}
\begin{figure}
\centering
\includegraphics[width=0.5\textwidth]{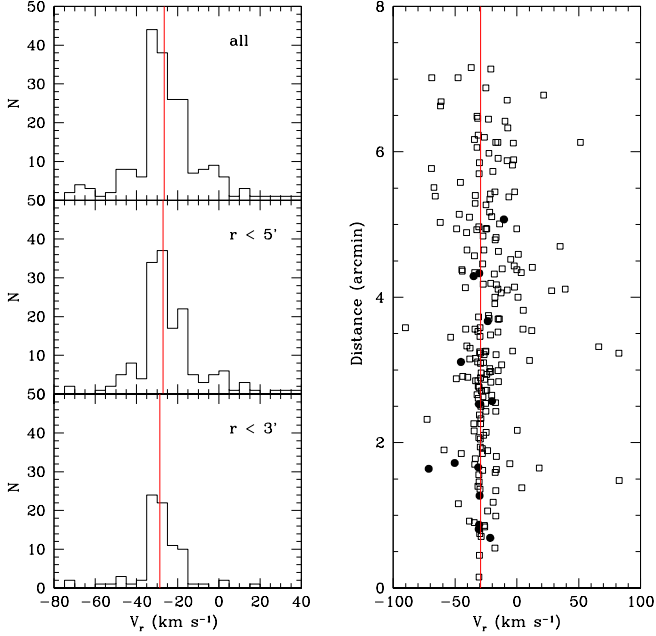}
\caption{Distribution of radial velocities for stars observed in the field of NGC 4815.  {\it  Left Panel:} Radial velocity distribution for  targets by distance from the cluster center, with the bottom panel showing stars within 3$\arcmin$, the middle panel showing stars within 5$\arcmin$, and the top panel showing the full sample.  The solid red line gives the average velocity computed for the distance range plotted in each panel, as explained in the text.  {\it Right panel:} Plot of the distance from the cluster center against radial velocity for all targets.  The solid black dots correspond to UVES observations.  The open squares correspond to GIRAFFE observations. }
\label{vr_histogram}
\end{figure}

As noted by all authors of photometric studies of NGC 4815, the cluster suffers from significant field star contamination.  As a result, radial velocities are critical in helping to determine cluster membership in any analysis of cluster properties.   Fig.~\ref{vr_histogram} shows the distribution of radial velocities measured from Gaia-ESO Survey observations of NGC 4815 using results from both GIRAFFE and UVES spectra.  Typical velocity errors for UVES spectra are 0.6 \kms, while for the GIRAFFE spectra errors range from 0.2 \kms\ to many tens of \kms\ for rapidly rotating stars.   The median velocity error for the GIRAFFE sample is 1.2 \kms.   The right panel in Fig.~\ref{vr_histogram} shows the full sample of objects, except for six stars outside the limits of the plot.  The left panels of the figure limit the range of velocities to within $\sim$ 60 \kms\ of the centroid of the distribution for increasing ranges of distance from the cluster center.  Open clusters are generally observed to have very small velocity dispersions, of 1-2 \kms\ \citep{merm09}, and the extremely broad velocity distribution displayed in the upper left panel of Fig.~\ref{vr_histogram} suggests it is dominated by field contamination.   One might expect that the cluster members would predominate over the field stars at distances closer to the center of the cluster.  And indeed, in Fig.~\ref{vr_histogram}, it appears that the distribution in velocities is less broad at the smaller distances, with a stronger sign of the cluster systemic velocity.  
Fig.~\ref{vr_histogram} shows that, in addition to the broad peak from $\sim -20$ to $-35$  \kms\ and the increasing dispersion with distance from the cluster center, the distribution  becomes increasingly asymmetric at larger distances, and shows an overall shift of the distribution to higher velocities.  

\begin{figure}
\centering
\includegraphics[width=0.5\textwidth]{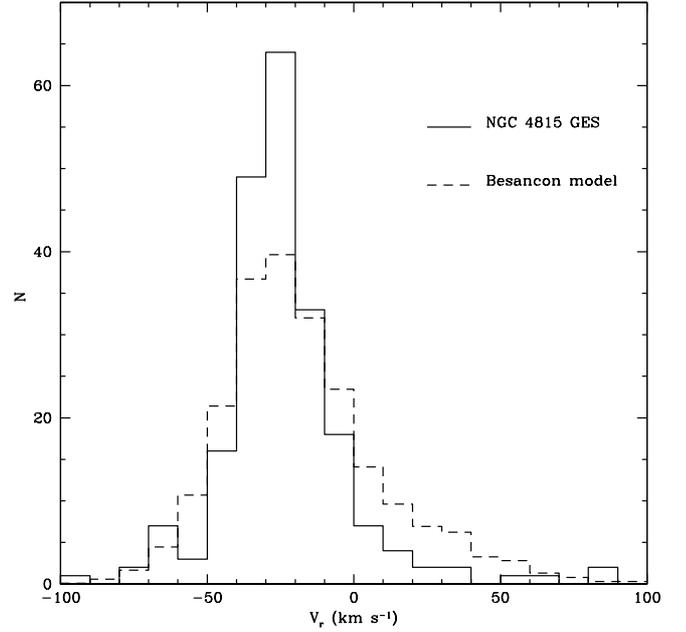}
\caption{Distribution of radial velocities for stars in the field of NGC 4815.  The solid line gives the observed radial velocities; the dashed line shows the  radial velocity distribution for stars in the magnitude and color ranges sampled by the NGC 4815 observations as predicted by the Besan\c{c}on model scaled to match the number of observed Gaia-ESO Survey targets. }
\label{vr_model}
\end{figure}

We can gain some insight into the impact of field star contamination on the cluster observations by looking at the expected velocity distribution for Galactic field stars according to the Besan\c{c}on model of stellar populations in the Milky Way \citep{besancon03}.  Utilizing the web-based version of the model (http://model.obs-besancon.fr), we have generated the anticipated velocity distribution for field stars that meet the Gaia-ESO Survey selection criteria on color and magnitude at the cluster's Galactic coordinates.    The resulting velocity distribution is shown in Fig.~\ref{vr_model} scaled to the number of observed targets, and superimposed over the Gaia-ESO Survey measured velocities.  The broad distribution seen in the Gaia-ESO Survey observations is clearly reflective of the expected Galactic field population, and explains the minor secondary peak and asymmetric tail to velocities higher than the cluster mean.  Based on this comparison to Besan\c{c}on models, we can see a clear excess of stars in the range of $-40$ to $-20$ \kms~ corresponding to the cluster velocity.    

The comparison with the Besan\c{c}on model results also suggests that we rely on the median of the distribution as a more robust measure of the cluster velocity to minimize the influence of the field star population that appears at higher velocities and to rely on the sample closest to the cluster center where the ratio of cluster members to non-members will be higher.   Limiting ourselves to consideration of the stars within 3 $\arcmin$ of the cluster center we find a median radial velocity of $-29.4$ \kms.  Applying an iterative 2-sigma clipping procedure on the median, as was done in the analysis of Tr 20 \citep{pdTr20}, results in the average velocities shown in each of  the distance ranges in Fig.~\ref{vr_histogram}.  This average velocity shifts to higher values with increasing distance from the cluster center,  as the larger field star contamination more strongly affects the average value.  While this shift is not large, from $-28.7$ \kms~ for distances inside 3$\arcmin$ to $-26.6$ \kms~ over the full 7.2$\arcmin$ field, we consider the velocity at the inner distances to be more reflective of the true cluster velocity.  As a result, we adopt the median value of $-29.4$ \kms~ as the cluster systemic velocity.  To obtain a representative dispersion about this value we rely again on the inner sample, which yields an rms deviation about the mean of 4 \kms, which reflects a combination of the intrinsic cluster velocity dispersion, observational measurement errors, the influence of undetected binaries, and the contamination by field stars.  

If we adopt a 1-sigma limit of 4 \kms~ about the cluster velocity of $-29.4$ \kms~ to identify the most likely cluster members, a total of 63 stars meet our criterion of membership over the entire range of distances sampled.  These numbers imply that only 29\% of the 218 observed stars have velocities consistent with membership within the cluster, although this sample very likely includes some field stars as well.  Scaling the results from the Besan\c{c}on model suggests that as many as half of the candidate members in the 1-sigma velocity range can be expected to be field stars.   The majority of field stars will be main-sequence dwarfs, and so will predominantly affect the GIRAFFE observations. 

The location of Gaia-ESO Survey targets  within this 1-sigma range are indicated in the lower panel of the color-magnitude diagram in Fig.~\ref{cmd_vi}.   Among the UVES targets, the radial velocity members cluster in the region of the red giant clump, as expected, but also include a brighter, cooler red giant in the sample.  The radial velocity members among the GIRAFFE observations do not follow a distinctive distribution along the main sequence, perhaps not surprising considering that there are expected to be an appreciable number of field stars with velocities similar to the cluster stars.  

\subsection{Stellar abundance analysis}
The analysis of the UVES spectra for NGC 4815 stars were carried out by the Gaia-ESO Survey working group WG11, whose work is described in Smiljanic et al. (2014, in prep.).  Atmospheric parameters and abundances were determined by up to 10 of the 14 independent WG11 nodes, using differing methods, but with the adoption of common model atmospheres (MARCS, \citealt{marcs08}), atomic parameters (Heiter et al. 2014, in prep.) and solar abundances \citep{grev07}.  Results from individual nodes are combined to yield the recommended atmospheric parameters, overall metallicity, and elemental abundances, as described by Smiljanic et al. (2014, in prep).   Here we discuss the recommended abundances for Fe and the Fe-peak elements Cr and Ni, the $\alpha$-elements Mg, Ca, Si, and Ti, and light elements Al and Na.  Elements that are affected by hyperfine splitting, which was not treated fully in the first Gaia-ESO Survey analysis, and neutron-capture elements will be analyzed in the future.  

Table~\ref{table_uves}  summarizes the recommended stellar and atmospheric parameters for UVES targets, as determined by the Gaia-ESO Survey spectral abundance analysis and reported in GESviDR1Final.  Errors on [Fe/H] and the atmospheric parameters are the rms dispersions  about the parameters from the individual analysis nodes.  GESviDR1Final also reported the mean of the errors on atmospheric parameters as provided by the individual nodes and these errors are larger than the scatter about the mean recommended values, usually  by about a factor of two. 

Of the 14 UVES targets, eight are non-members based on their radial velocities, where we use the criterion of falling more than 1-sigma away from the cluster's systemic velocity of $-29.4$ \kms. Of the six candidate members based on radial velocity, one of them could not be analyzed for abundances because of its very broad-lined nature (star \#725, GES ID 12574080-6455572).  The remaining five stars have a mean $<$[Fe/H]$> =+0.03 \pm 0.05$ (s.d.), which we adopt as the cluster metallicity for the remaining analysis.  These stars have a mean velocity of $<V_{r}> = -30.6 \pm 0.4$ \kms, showing very little dispersion.  We note that this velocity is lower by 1.2 \kms~  relative to the cluster mean determined largely by the more numerous GIRAFFE observations.  This  offset between the two data sets is slightly larger than that found by \citet{pdTr20} for Tr 20, but very similar to that found by \citet{g2v} from observations of Gamma Velorum.  Only one star in the field of NGC 4815 was observed with both UVES and GIRAFFE, but it, too, shows a velocity offset of 1 \kms~ with the UVES velocity being lower.   The source of this discrepancy is not known and is under investigation.

\begin{table*}
\caption{Parameters for stars in the field of NGC 4815 with UVES observations}
\begin{center}
\scriptsize
\begin{tabular}{rcrrrccccccc}
\hline\hline
ID & GES ID & $V$ & $B-V$ & $V-I$ & RA & Dec & $V_{r}$ &  [Fe/H] & $T_ {\rm eff}$ & log $g$  & $\xi$   \\
     &               &     &         &       &    &         &  (\kms)   &   &   (K)       &             &  (\kms)    \\  
\hline\hline
  76   &   12575308-6457182 & 14.429 & 1.383  & 1.763  & 194.4711667 & -64.95505556  &   $-21.60$ &  $-$0.42$\pm$0.17 &  5172$\pm$ 227 &  2.97$\pm$0.76 &  0.65$\pm$0.55   \\
    95   &   12575529-6456536 & 14.128 & 1.590  & 1.730  &  194.4803750  & -64.94822222   & $-30.85$ &  $-$0.03$\pm$0.09 &  5068$\pm$73 &   2.79$\pm$0.25 &  1.15$\pm$0.12 \\
  106  &   12580262-6456492 & 14.158 &  1.613  & 1.896 &  194.5109167 & -64.94700000     &    $-30.56$ &   +0.11$\pm$0.06 &  4926$\pm$77 &   2.57$\pm$0.08  & 1.53$\pm$ 0.14 \\
  210  &   12575511-6458483 & 14.043 & 1.649  & 1.841  & 194.4796250  & -64.98008333  &  -$30.16$  &  +0.06$\pm$0.07&   4870$\pm$84   & 2.55$\pm$0.47 &  1.43$\pm$0.24 \\
  341  &   12574905-6458511 & 14.709&    & 1.783  & 194.4543750  & -64.98086111  &  $-71.32$  &  $-$0.15$\pm$0.20 &  4760$\pm$120 &  2.56$\pm$0.25  & 1.40$\pm$0.17 \\
 358  &   12574328-6457386 & 13.703 & 1.702  & 1.946  & 194.4303333 & -64.96072222  &  $-31.27$ &  +0.00$\pm$0.08 &  4895$\pm$40 &   2.40$\pm$0.13  &  1.73$\pm$ 0.14  \\
  374   &  12574341-6458045 & 14.564 & 1.544  & 1.805  & 194.4308750  & -64.96791667  &  $-50.33$  & $-$0.06$\pm$0.05 &  4950$\pm$50  &  2.74$\pm$0.23 &  1.39$\pm$0.09  \\
  725    &  12574080-6455572 & 14.186 & 1.495  & 1.769 &  194.4200000     &  -64.93255556  &  $-30.72$ &   & & &   \\
  743   &  12574457-6459398 & 14.376 & 1.419  & 1.581  & 194.4357083 & -64.99438889  &  $-20.17$  & +0.06$\pm$0.07 &  4658$\pm$95  &  3.00$\pm$0.36  &  1.18$\pm$ 0.10 \\
  1024  &  12575531-6500412 & 14.015 & & 1.541 &  194.4804583 &  -65.01144444  &  $-45.27$  & $-$0.13$\pm$0.11 &  4823$\pm$ 87 &   2.79$\pm$0.19  & 1.23$\pm$0.03 \\
  1355  & 12573217-6455167 & 14.190  & 1.609  & 1.820  &  194.3840417 & -64.92130556  &  $-23.57$  &  +0.05$\pm$0.05 &  5048$\pm$102 &  2.60$\pm$0.18 &  1.47$\pm$0.14  \\
  1763 &   12581939-6453533 & 14.874 & 2.310 &  1.660  &  194.5807917 & -64.89813889  &  $-34.99$ &  $-$0.14$\pm$0.19 &  4032$\pm$52  &  1.42$\pm$0.24 &  1.61$\pm$0.31 \\
    1795  & 12572442-6455173 & 13.482 & 1.891 &  2.273 & 194.3517500 &   -64.92147222  &  $-30.41$ &   +0.03$\pm$0.07 &  4198$\pm$88  &  1.56$\pm$0.37 &  1.37$\pm$0.05  \\
    2356  &  12571312-6456090 & 14.811 & 1.671 &    &  194.3046667 & -64.93583333  &  $-10.52$  &  $-$0.04$\pm$0.11 & 4745$\pm$92 &  2.60$\pm$0.23  & 1.39$\pm$0.08  \\
      \hline
\hline
\end{tabular}
\label{table_uves}
\end{center}
\end{table*}

\begin{table*}
\caption{Abundances for stars in the field of NGC 4815}
\begin{center}
\scriptsize
\begin{tabular}{lcccccccccc}
\hline\hline
ID & member? & [Fe/H]  & [Na/Fe] & [Al/Fe] & [Mg/Fe] & [Si/Fe] & [Ca/Fe] &  [Ti/Fe] & [Cr/Fe] & [Ni/Fe]   \\
\hline\hline
   95   &  y &  $-$0.03$\pm$0.09 & +0.31$\pm$0.04 & +0.13$\pm$0.01 &  +0.03$\pm$0.16& $-$0.06$\pm$0.05 & $-$0.06$\pm$0.03 & $-$0.10$\pm$0.02 & $-$0.17$\pm$0.07& $-$0.09$\pm$0.05  \\
  106  &  y & +0.11$\pm$0.06 &  +0.26$\pm$0.01 &  +0.05$\pm$0.02  & +0.05$\pm$0.17& +0.00$\pm$0.04 & $-$0.05$\pm$0.05 & $-$0.10$\pm$0.02& $-$0.15$\pm$0.05 & $-$0.11$\pm$0.01 \\
  210  &  y & +0.06$\pm$0.07&   +0.31$\pm$0.04   & +0.07$\pm$0.01 &  +0.19$\pm$0.07& $-$0.07$\pm$0.01 & $-$0.05$\pm$0.02 & $-$0.16$\pm$0.03& $-$0.14$\pm$0.07 & $-$0.11$\pm$0.08 \\
 358  & y &  +0.00$\pm$0.08 & +0.40$\pm$0.11 &  +0.11$\pm$0.03  &  +0.15$\pm$0.10 & $-$0.04$\pm$0.05 & $-$0.02$\pm$0.05 & $-$0.15$\pm$0.03 & $-$0.10$\pm$0.03 & $-$0.10$\pm$0.04 \\
   1795  &  y & +0.03$\pm$0.07 &  +0.28$\pm$0.03  &  +0.06$\pm$0.01 &  +0.29$\pm$0.04 & +0.06$\pm$0.10 & $-$0.11$\pm$0.07 & $-$0.27$\pm$0.03 & $-$0.17$\pm$0.01 & $-$0.06$\pm$0.04 \\
   \hline
Mean (all) &  &  +0.03$\pm$0.05 & +0.31$\pm$0.05 & +0.08$\pm$0.03 & +0.14$\pm$0.11 & $-$0.02$\pm$0.05 & $-$0.06$\pm$0.03 & $-$0.16$\pm$0.07 & $-$0.15$\pm$0.03 & $-$0.09$\pm$0.02 \\
Mean (- 1795) &  & +0.03$\pm$0.06 & +0.32$\pm$0.06 & +0.09$\pm$0.04 & +0.11$\pm$0.08 & $-$0.04$\pm$0.03 & $-$0.04$\pm$0.02 & $-$0.13$\pm$0.03 & $-$0.14$\pm$0.03 & $-$0.10$\pm$0.01 \\
\hline 
  & & & & & & & & & & \\
  76   &   n &  $-$0.42$\pm$0.17 &  +0.41$\pm$0.17 &  +0.35$\pm$0.11 &  +0.01$\pm$0.08  & $-$0.18$\pm$0.05 & +0.04$\pm$0.06& +0.11$\pm$0.02 & $-$0.01$\pm$0.13  & $-$0.05$\pm$0.06 \\ 
  341  & n &  $-$0.15$\pm$0.20 &  +0.14$\pm$0.14 &  +0.18$\pm$0.02  & +0.20$\pm$0.11& +0.00$\pm$0.03 & $-$0.04$\pm$0.04 & $-$0.04$\pm$0.04 & $-$0.17$\pm$0.11 & $-$0.02$\pm$0.05 \\
  374   & n &  $-$0.06$\pm$0.05 &  +0.09$\pm$0.04 &  +0.11$\pm$0.02 &  +0.13$\pm$0.07 & $-$0.07$\pm$0.03 & +0.00$\pm$0.02 & $-$0.01$\pm$0.03 & $-$0.10$\pm$0.02& $-$0.13$\pm$0.02  \\
  743   &  n & +0.06$\pm$0.07 &  +0.19$\pm$0.06  &  +0.07$\pm$0.01  &  +0.21$\pm$0.08& +0.01$\pm$0.06  & $-$0.09$\pm$0.02& $-$0.15$\pm$0.02  & $-$0.20$\pm$0.06 & +0.01$\pm$0.04 \\
 1024  & n &  $-$0.13$\pm$0.11 &  +0.12$\pm$0.05 &  +0.17$\pm$0.01  & +0.15$\pm$0.04& $-$0.02$\pm$0.02& $-$0.02$\pm$0.02 & $-$0.01$\pm$0.01 & $-$0.11$\pm$0.03 & $-$0.10$\pm$0.02 \\
 1355  &  n? & +0.05$\pm$0.05 &  +0.32$\pm$0.06 & +0.10$\pm$0.03 &  +0.10$\pm$0.12 & $-$0.06$\pm$0.05 & $-$0.03$\pm$0.04 & $-$0.09$\pm$0.03 & $-$0.08$\pm$0.04 & $-$0.13$\pm$0.02 \\
 1763 &  n &  $-$0.14$\pm$0.19 &  +0.09$\pm$0.01   &  +0.40$\pm$0.01 &  +0.50$\pm$0.03& +0.10$\pm$0.14& +0.04$\pm$0.09 & +0.01$\pm$0.02 & $-$0.09$\pm$0.10 & +0.07$\pm$0.03 \\
  2356  &  n & $-$0.04$\pm$0.11 & +0.04$\pm$0.06 &  +0.14$\pm$0.02  & +0.13$\pm$0.12 & +0.00$\pm$0.04& $-$0.07$\pm$0.03  & $-$0.04$\pm$0.03 & $-$0.20$\pm$0.06 & $-$0.06$\pm$0.04 \\
  \hline
Mean & & $-$0.11$\pm$0.16 & +0.19$\pm$0.12 & +0.20$\pm$0.13 & +0.19$\pm$0.15 & $-$0.03$\pm$0.09 & $-$0.01$\pm$0.05 & $-$0.03$\pm$0.08 & $-$0.11$\pm$0.06 & $-$0.05$\pm$0.08 \\

      \hline
\hline
\end{tabular}
\label{table_abund}
\end{center}
\end{table*}

Tab.~\ref{table_abund} summarizes the abundances resulting from the Gaia-ESO Survey analysis of the UVES observations, distinguishing in separate sections the stars considered as secure members from those with velocities that fall more than 4 \kms~ from the adopted cluster velocity.  The elemental abundance ratios presented are calculated from the [Fe/H] values determined for each star individually and are based on the \citet{grev07} solar abundances.  The errors of the [X/Fe] ratios in Tab.~\ref{table_abund}  are those reported for each element in the internal data release, and do not include the error associated with the [Fe/H] value nor the propagation of errors in the atmospheric parameters.   Below each of the groupings of members and non-members the averages across stars in each category are given; the errors listed are standard deviations about the mean abundance.  

GIRAFFE spectra taken with grating HR15n were analyzed for atmospheric parameters and abundances by WG10, whose work is described in Recio-Blanco et al. (2014, in prep.).  The limited number of lines available for analysis, the generally lower S/N, and the broad-lined nature of many of the spectra prevented the analysis for many stars.  Metallicities were obtained for only 25 stars, and of these, only 3 are members according to the 1-sigma criterion for radial velocities applied above, while two more fall within 2-sigma of the cluster systemic velocity.  These stars give [Fe/H] values of 0.21 to 0.4, substantially higher than the values for the cluster members from the UVES observations.   The one star observed with both spectrographs (\#2356, GES ID 12571312-6456090, a velocity non-member), was found to have a metallicity of $+0.15 \pm 0.18$ from the GIRAFFE observation, also suggesting an offset of $\sim$ 0.2 dex between the metallicities derived from the GIRAFFE and the UVES spectra.  The cause of this systematic difference between abundance scales is still not fully understood and under study.  For the reminder of this paper we will rely on the results from the higher-resolution UVES spectra.

Considering the UVES abundances for the five members of NGC 4815, we can see from  Tab.~\ref{table_abund} generally very good agreement in abundance ratios.  For almost all elements, the standard deviations about the mean for cluster members are less than or comparable to that found for iron, at 0.05 dex.  The dispersion is somewhat higher only for [Mg/Fe], at  0.1 dex.   However, the errors on the abundance determinations for individual stars are typically somewhat higher for Mg than the other elements, likely contributing to the larger scatter.    

Among the radial velocity members of NGC 4815, four of them are red giant clump stars, with very similar temperatures and gravities.  Star  \#1795, GES ID 12572442-6455173, is a much cooler and more luminous star on the red giant branch, with a temperature of 4200K.  As a result, many lines are very strong, may be blended and are more challenging to measure.  This fact might explain the somewhat discrepant abundances in Mg and Ti seen in \#1795 compared to the cluster mean.   Omitting star 1795 from the abundance averages  reduces the  dispersion about the mean in almost all cases, resulting in dispersions of only 0.03 dex for most elements, again with the exception of Mg.  

The small dispersions in abundance are consistent with those found in other studies of the chemical homogeneity of open clusters \citep{deSilva06, deSilva07}.  Such homogeneity is not surprising for a cluster of the rather low mass of NGC 4815, estimated to be $880 \pm 230  M_{\sun}$ for stars more massive than $\sim 0.8 M_{\sun}$ \citep{pris01}.  \citet{BKF2010} show that open clusters are expected to be uniform for their observed densities for masses up to $10^{5} M_{\sun}$.  While NGC 4815 may have lost mass in its initial collapse and subsequent evolution, it is unlikely to have been initially above this limit.

These final abundance ratios for the cluster show a substantial range in value.  Among the $\alpha$-elements, we see signs of modest enhancement in [Mg/Fe] values, but  solar values of [Si/Fe] and [Ca/Fe].   [Ti/Fe] appears slightly depleted.  This behavior is similar to that found in other samples of open clusters, supporting the observation that the $\alpha$-elements arise from multiple nucleosynthetic sources (e.g. \citealt{yong12,thiel86}).  Turning to the Fe-peak elements [Cr/Fe] and [Ni/Fe], we see slightly sub-solar ratios for both elements with very small internal dispersions.  We recall, however, that these abundances are differential with respect to the \citet{grev07} solar abundances, and not with respect to solar abundances derived in the Gaia-ESO Survey, so systematic offsets in the overall abundance scale for these, and all abundance ratios, is a possibility.  

The light element ratio [Na/Fe] is significantly and uniformly enhanced in NGC 4815 stars, at $<$[Na/Fe]$> = +0.31 \pm 0.05$.    [Al/Fe] shows a more modest, but well defined enhancement of $<$[Al/Fe]$> = +0.08 \pm 0.03$.   These values are generally consistent with what is found in evolved stars in open clusters (Jacobson et al. 2011).   These enhancements may be due to a variety of causes, from the adoption of the approximation of LTE in the abundance analysis to the evolutionary effects of internal mixing and the dredge up of processed material to the atmospheres of these evolved stars \citep{sm12}.  The complex situation regarding Na abundances and their analysis in Gaia-ESO Survey cluster data will be addressed in future work.  

Overall, this abundance profile along with the very small internal abundance dispersion among the NGC 4815 radial velocity members, provides a means to chemically tag the cluster members.   Our limitation in velocity to define members was a rather strict one, taken more to ensure we were not contaminated by non-members than to be inclusive of possible members.   As a result, it is very possible that some of the stars in this non-member sample with velocities within $\sim$ 10 \kms~ of the cluster mean are indeed binary stars that are members of the cluster.  In fact, of the UVES observations, star \#76, GES ID 12575308-6457182, was identified as a spectroscopic binary in the reduction and analysis process; its velocity of $-21$ \kms~ differs by less than 10 \kms~ from the mean.  Its metallicity, [Fe/H] $=-0.42 \pm 0.17$, however, indicates it is not a cluster member.   In this context it is interesting to look at the behavior of the stars we have denoted as velocity non-members observed in the field of NGC 4815. 

\begin{figure}
\centering
\includegraphics[width=0.5\textwidth]{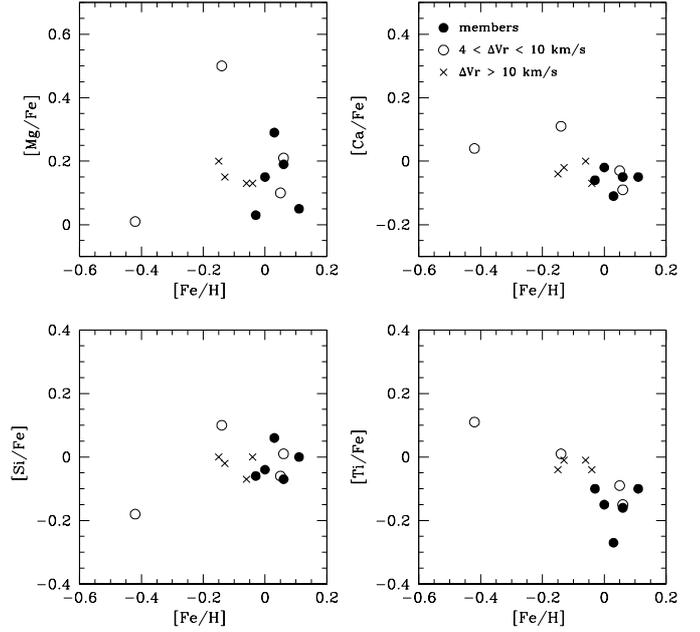}
\caption{Distribution of $\alpha$-element abundances with [Fe/H] for stars observed in the field of NGC4815.  Stars with radial velocities within 4 \kms~ of the cluster mean are considered members and denoted with solid circles; stars with velocities that deviate by up to 10 \kms~ are denoted by open circles; stars with velocities more then 10 \kms~ from the cluster mean are denoted by  crosses. }
\label{alpha_abund}
\end{figure}

\begin{figure}
\centering
\includegraphics[width=0.5\textwidth]{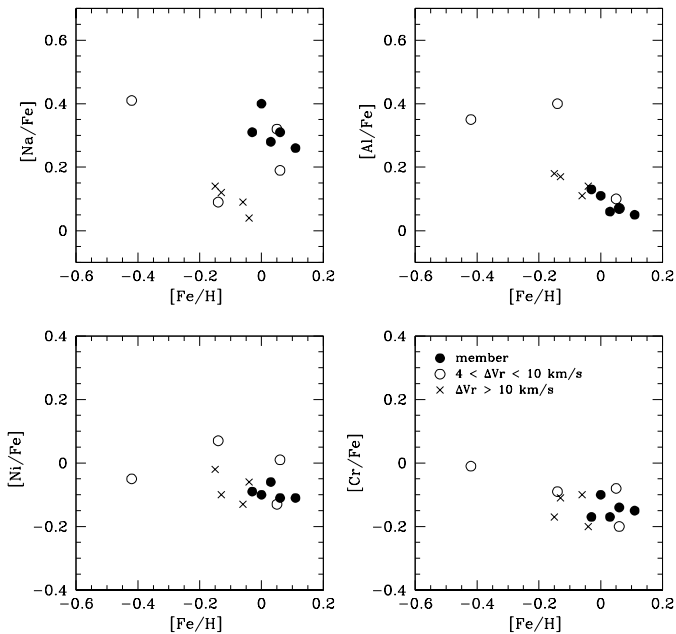}
\caption{Distribution of the abundances of light elements Na and Al and Fe-peak elements Ni and Cr against [Fe/H] for stars observed in the field of NGC4815.  Symbols are as in previous figure. }
\label{other_elem}
\end{figure}

The lower panel of Tab.~\ref{table_abund} gives the recommended abundances for the stars with velocities more than 4 \kms~ from the cluster mean; the last line in the table shows the averages in abundances across this sample.   Not surprisingly,  the dispersions about the mean values for these stars show significantly higher values, ranging up to 0.15 dex, typically several times that found for the stars at the cluster velocity.  The distribution of abundances among all the NGC 4815 stars measured are shown in Fig.~\ref{alpha_abund} for the $\alpha$-elements, and in Fig.~\ref{other_elem} for the remaining elements.  Points are coded by velocity range with divisions made at deviations of within 1-sigma, 1 to 2.5-sigma, and more than 2.5-sigma from the cluster mean.  

Considering the $\alpha$-elements, stars that overlap with the cluster [Fe/H] value, regardless of their radial velocity, appear to fall within the range seen in the cluster members.  In particular, the two stars with velocities only 10 \kms~ from the cluster mean have $\alpha$-element abundances that are consistent with those of NGC 4815.  Consideration of the behavior in other elements indicates that star \#743, GES ID 12574457-6459398, deviates from the cluster abundance pattern notably in [Ni/Fe] (where it is higher by 0.1 dex) and  [Na/Fe] (where it is lower by 0.1 dex).  However, star \#1355, GES ID 12573217-6455167, has values in all  abundances that are consistent with membership.  With a velocity that is within 2-sigma of the cluster mean velocity, star \#1355 might be best denoted a possible cluster member.  

Although the number of points is limited, the overall trends of elemental abundances with [Fe/H] in  Fig.~\ref{alpha_abund} and Fig.~\ref{other_elem}, showing a slight increase in abundance ratio with decreasing [Fe/H] and a substantial scatter in the field population, are consistent with trends seen in larger samples of disk stars.   The uniformity of both the observational data and the abundance analysis of the Gaia-ESO Survey data offers new insight into the differences detectable in the cluster and field populations and we refer the reader to a first analysis of the chemical tagging of Gaia-ESO Survey open clusters, including NGC 4815, in \citet{magrini14}.

\section{Isochrone fitting}
The estimate of the cluster parameters, namely age, distance, and reddening, is obtained by means of the isochrone fitting method. We have adopted three different sets of isochrones to have a less model-dependent solution: the PARSEC  \citep{parsec}, BASTI \citep{basti}, and  Dartmouth \citep{dartmouth} 
models.
In principle, the best-fitting isochrone is chosen as the one which can describe at the same time the main age-sensitive evolutionary phases: the luminosity and color of the main sequence turn-off (MSTO), red hook (RH), and red clump (RC) when possible.  As we will see below, in the case of NGC 4815, a simultaneously good fit of all of these features in published photometry is not possible.   We used the metallicity obtained from the UVES abundance analysis, as discussed above, i.e. solar metallicity.

The errors on the estimated parameters are mainly due to the uncertainties in the delineation of the age indicators in the CMD. In particular, the field contamination complicates the correct definition of the RC position. The membership information for the RC stars is then especially important for a robust interpretation of the fit. The internal photometric errors do not have a significant impact on the error budget. 

\begin{figure}
\centering
\includegraphics[scale=0.45]{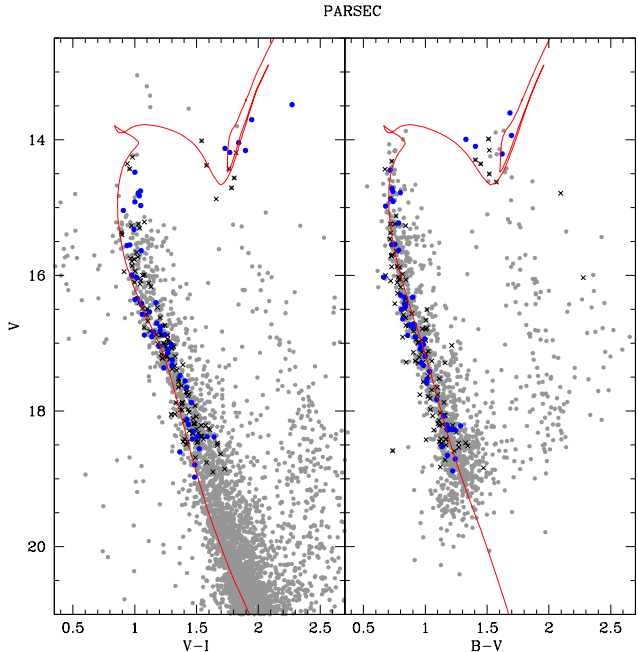} 
\caption{The CMDs are obtained for stars inside 3$\arcmin$. The blue points are all the Gaia-ESO Survey targets candidate members while the black crosses are all the non-members.  The continuous line is the best fit isochrone using the PARSEC isochrones attempting a compromise fit to both photometric sets as described in the text.  See Table \ref{tab:isoc} for the adopted parameters for the isochrone fitting.} 
\label{cmd_par}
\end{figure}

In order to obtain an accurate estimate we make use of both the $V,B-V$ \citep{co94} and 
$V,V-I$ \citep{pris01} CMDs, and of the membership of the target stars described in Sect. 3.2. These data sets, plotted in Fig.~\ref{cmd_par} for stars within 3\arcmin~ of the cluster center, show that the observational CMD in the $V,B-V$ plane provides a clearer description of the bright part of the MS, especially near the MSTO.   The MS in the $V,B-V$ diagram appears to reach to brighter magnitudes and is better defined than in the $V,V-I$  diagram.  This behavior is confirmed in the earlier $BV$ photometry of \citet{kf91}, giving us confidence that the MS in these inner regions, where the cluster sequences are more pronounced, extends to $V\sim 14$.  We note that the Prisinzano et al. photometry was intended to sample the mass function of the cluster by working to faint magnitudes.  In fact, the $VI$ photometry shows a number of bright stars with surprisingly large errors, suggesting that the $VI$ photometry is incomplete and subject to large errors due perhaps to saturation effects at the brightest magnitudes.  As a result, we rely only on the fainter sequence in the $V,V-I$ CMD, where the detailed shape of the MS is a useful constraint for the estimate of cluster parameters.  

We find that the best fitting solution for one observational plane is not perfect for the other when a standard extinction law is considered ($R_V=3.1$ and $E(V-I)=1.25\times E(B-V)$, see \citealt{dean78}), although the mismatch is small.   This failure in simultaneously fitting the CMDs in two colors has been found in other cases (see e.g., \citealt{ahu13,pdTr20}) but no definitive explanation has been found. In addition to the possibility of zero point errors in the photometry, there is the well known issue of the photometric transformations from the theoretical to the observational plane. The three models use different transformations, adding a source of uncertainty in these comparisons, but also allowing an assessment of the limitations to the derived parameters.

We have attempted to derive a best fit that considers features in both diagrams that give the maximum age discrimination while giving more weight to the $BV$ photometry for the reasons given above. 
In Fig.~\ref{cmd_par} we show the best fit for the PARSEC models. The blue points are all the target members based on radial velocities while the black crosses are all the target non-members, regardless of distance from the cluster center.   In deciding on the best fit isochrone, we emphasized the fit to the upper MS and turnoff region in the $V,B-V$ diagram, where the match is quite good, even though the fit departs slightly from the observed sequence at fainter magnitudes.   In the $V,V-I$ diagram, we instead utilized stars with $V \sim 16 - 19$ to define the MS shape, where the deflection in the theoretical isochrones is a good age discriminant.  In the $V,V-I$ diagram, the upper MS and turnoff region is not well defined and is not well fit by the isochrones, as anticipated.  By relying on the definition of the MSTO in  $V,B-V$, we find the $V-I$ data in general are slightly too red relative to the isochrone.   Our final adopted parameters, given in  Tab.~\ref{tab:isoc}, are then a compromise that does not perfectly fit either set of photometry  but utilizes features of both.   With the PARSEC isochrones, we find an age of $0.63 \pm 0.1$ Gyr,  reddening $E(B-V) = 0.65 \pm 0.02$, and a distance modulus of $(m-M)_0 = 12.20 \pm 0.1$.  Errors given on the fitted parameters take into account the additional uncertainty in trying to match both photometric systems.  

\begin{figure}
\centering
\includegraphics[scale=0.45]{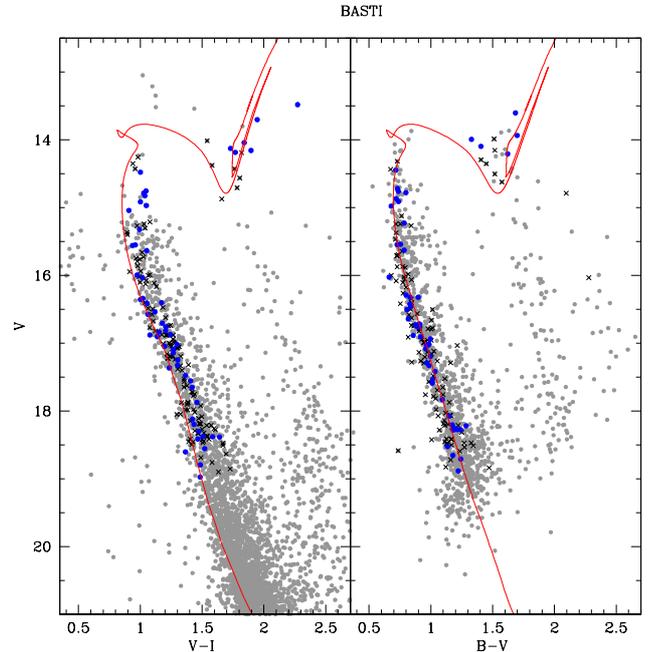} 
\caption{As in Fig.~\ref{cmd_par}, but with fits to the BASTI isochrones.} 
\label{cmd_bas}
\end{figure}

\begin{figure}
\centering
\includegraphics[scale=0.45]{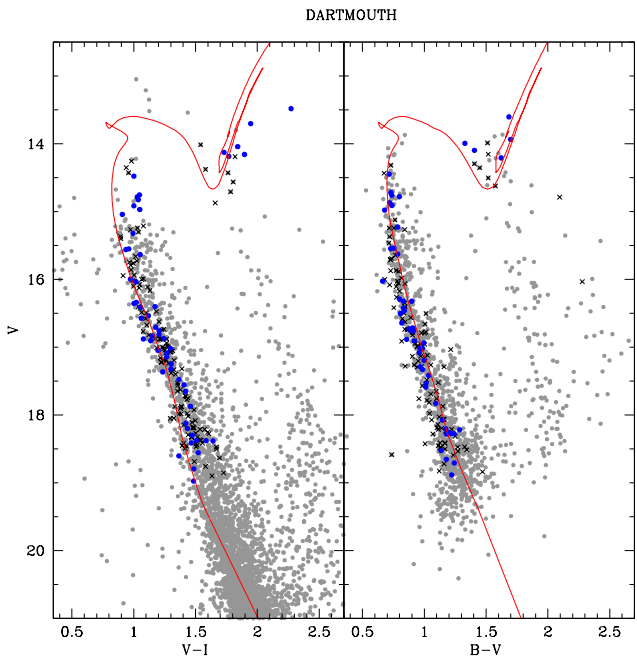} 
\caption{As in Fig.~\ref{cmd_par}, but with fits to the Dartmouth isochrones.} 
\label{cmd_dart}
\end{figure}

These best fitting isochrones describe quite well the luminosity of the RC and the MSTO (defined in $BV$) and follow the shape of the MS, while the RC color is slightly redder than observed.  
The fits to the BASTI and Dartmouth isochrones are shown in Fig.~\ref{cmd_bas} and Fig.~\ref{cmd_dart}, respectively, and given in Tab.~\ref{tab:isoc}; 
 the same considerations hold also for these models.   We note that all models produce consistent determinations of cluster parameters, within the uncertainties, yielding an age of 0.5 to 0.63 Gyr with a turnoff mass of 2.5 - 2.7  $M_{\sun}$, and a distance of 2.5 - 2.7 kpc.  These parameters are generally consistent with previous estimates, but with slightly larger ages on average and somewhat lower reddening.  

The relatively high reddening found from these fits, of $E(B-V) \sim 0.6$, along with the low latitude of the cluster suggests that the cluster field may suffer from variable reddening.  The very high contamination by field stars, coupled with the relatively low mass of NGC 4815, makes a detailed derivation of the differential reddening across the field, as was done for Tr 20 \citep{pdTr20},  problematic.  However, in an effort to define the magnitude of differential reddening, we have carried out a similar analysis, finding a range of approximately 0.1 mag in $E(B-V)$ across the cluster field of view.  A variability in reddening of this magnitude is not surprising, and contributes to the width of the main sequence in Fig.~\ref{cmd_par} - Fig. \ref{cmd_dart}, even among the radial velocity members.  This potential variability in the reddening is accounted for in the parameters for the cluster presented in Tab. \ref{tab:isoc}.

\begin{table*}
\centering
\caption{ Results using different evolutionary models with solar metallicity. The $R_{GC,\odot}$ adopted is 8 kpc (see \citealt{mal13}).}
\begin{tabular}{lccccccc}
\hline
\hline
 Model & age & $(m-M)_0$ & $E(B-V)$ & $d_{\odot}$ & $R_{GC}$ & $Z$ & $M_{TO}$ \\
       & (Gyr) & (mag) & (mag) & (kpc) & (kpc) & (pc) & ($M_{\odot}$)\\
\hline
\hline
PARSEC   & 0.63$\pm$0.1 & 12.00$\pm$0.1 & 0.65$\pm$0.02 & 2.51 & 6.93 &  -92.0 & 2.5\\
BASTI    & 0.50$\pm$0.1 & 12.20$\pm$0.1 & 0.64$\pm$0.02 & 2.75 & 6.87 & -100.8 & 2.7\\
DARTMOUTH & 0.60$\pm$0.1 & 11.95$\pm$0.1 & 0.59$\pm$0.02 & 2.45 & 6.95 & -89.8 & 2.5\\
\hline
\end{tabular}
\label{tab:isoc}
\end{table*}

\section{Summary and Conclusions}
We have used the results from Gaia-ESO Survey spectroscopic data and published photometric data to carry out a study of the intermediate age cluster NGC 4815.   Observations of 218 stars in the field of the cluster provide radial velocities from which we estimate the cluster systemic velocity to be $-29.4$ \kms.  The radial velocity distribution is, however, quite broad, suggesting significant contribution by field stars, which is also indicated by the Besan\c{c}on Galactic stellar populations model.  Selecting stars within 4 \kms~ of this mean velocity as potential cluster members results in only $\sim$30\% of the observed Gaia-ESO Survey targets having velocities consistent with membership.  

Stellar abundances from the  UVES observations for the five evolved stars that are likely members yield a mean cluster metallicity of [Fe/H] = +0.03 $\pm$ 0.02 dex (error in the mean).  Elemental abundances also show a small dispersion about the mean values, indicating homogeneous chemical composition in the cluster.  Among the $\alpha$-elements, [Ca/Fe] and [Si/Fe] appear solar, while [Mg/Fe] appears somewhat enhanced and [Ti/Fe] slightly depleted.   Iron-peak elements [Ni/Fe] and [Cr/Fe] are slightly underabundant relative to solar.  Consistent with findings for many other open clusters, the light elements Na and Al are enhanced over solar, but what role the effect of inadequacies in the assumed LTE analysis or internal stellar mixing might have in explaining these abundances remains to be investigated.  Abundances for radial velocity non-members of NGC 4815, sampling the disk field population, show typically much larger dispersions, but have abundance distributions that generally overlap with the cluster values at the same metallicity, particularly for the $\alpha$-elements.  There are indications, once all elements are considered, that the detailed abundance pattern of the cluster is distinguished from those of the field stars.  Obtaining abundances for additional elements will help clarify this picture.  Future releases will include the determination of more elements; the neutron-capture elements will be especially interesting in this regard.  It is expected that as the survey proceeds, the internal precision of the GES abundances may be improved through analysis of an increased number of calibrating benchmark stars and a fine-tuning of the homogenization process leading to recommended parameters.  In addition, for many clusters sample sizes will be larger; as one of the less massive open clusters, the number of potential targets in NGC 4815 is limited.  However, it is clear that the uniformity of the Gaia-ESO Survey analysis and the homogeneity of the resultant abundances offers the opportunity to explore the potential systematic chemical differences between the cluster and the field at a new level of detail, for many open clusters, in many lines of sight.  These data combined with proper motions from Gaia will also enable a more thorough kinematic selection and dynamical investigation of the cluster and its surrounding field population. 

Finally, using the information on radial velocity membership, and with the cluster metallicity constrained by the Gaia-ESO Survey determinations, theoretical isochrones were fit to the published $BV$ and $VI$ photometry of NGC 4815 to derive more robust estimates of the cluster parameters.  Utilizing three different sets of isochrones and fitting both the $VI$ and $BV$ color-magnitude diagrams provided an assessment of uncertainty in these parameters.  We conclude that NGC 4815 has an age of between 0.5 to 0.63 Gyr, a reddening $E(B-V) = 0.59$ to 0.65, and distance modulus $(m-M)_0$ of 11.95 to 12.20, placing it at a distance of 2.45 to 2.75 kpc, or 6.9 kpc from the Galactic center.  The isochrones provide a good fit to the shape of the main-sequence and the luminosities of the turnoff and the red clump stars, but are less successful in matching simultaneously the $BV$ and $VI$ colors.   Possible reasons for this mismatch could rest in differences in the accuracy of the photometric zero points, or the difficulty of transforming the isochrones from the theoretical to the observational plane.

\begin{acknowledgements}
      This research has made use of the WEBDA database, originally developed by J-C. Mermilliod, now operated at the Department of Theoretical Physics and Astrophysics of the Masaryk University, and of the BaSTI web tools for stellar isochrones.  The results presented here benefitted from discussions in three Gaia-ESO workshops supported by the ESF (European Science Foundation) through the GREAT (Gaia Research for European Astronomy Training) Research Network Program (Science meetings 3855, 4127, and 4415).  We acknowledge the support from INAF and Ministero dell' Istruzione, dell' Universit\`a e della Ricerca (MIUR) in the
form of the grant ÒPremiale VLT 2012Ó. This work was partially supported by the Gaia Research for European Astronomy Training
(GREAT-ITN) Marie Curie network, funded through the European Union Seventh Framework Programme [FP7/2007-2013] under grant agreement no 264895.  It was also partly supported by the European Union FP7 programme through ERC grant number 320360 and by the Leverhulme Trust through grant RPG-2012-541.
      
\end{acknowledgements}


\end{document}